\documentclass[12pt]{article}

\usepackage{epsfig} 

\newcommand{\be}{\begin{equation}} 
\newcommand{\en}{\end{equation}}
\newcommand{\bea}{\begin{eqnarray}}
\newcommand{\ena}{\end{eqnarray}}

\newcommand{\hbo}{\hbox to 1 true cm {\hfill } } 
\newcommand{\tr}{\hbox{tr}}

\begin{document}

\vglue 1truecm
  
\vbox{ 
UNITU-THEP-31/02 \hfill December 17, 2002
}
  
\vfil
\centerline{\large\bf Lattice effective theory and the phase transition } 
\centerline{\large\bf at finite densities} 

\bigskip
\bigskip
\bigskip
\centerline{ Kurt Langfeld } 
\bigskip
\vspace{.5 true cm} 
\centerline{ Institut f\"ur Theoretische Physik, Universit\"at 
   T\"ubingen }
\centerline{ D-72072 T\"ubingen, Germany. }
\bigskip\bigskip

\vskip 1.5cm
\begin{abstract}
The transition from the hadronic phase to the phase of 
color-superconductivity at large densities is addressed by an
effective theory which incorporates the Yang-Mills 
dynamics in addition to the di-quark degree of freedom. A toy 
version of this theory is studied by lattice simulations. 
A first order phase transition separates the regime of 
broken color-electric flux tubes from the color superconducting 
phase. My findings suggest that the quark and gluon liberation 
occurs at the same critical chemical potential. 
\end{abstract}

\vfil
\hrule width 5truecm
\vskip .2truecm
\begin{quote} 
{\footnotesize
\centerline{\tt email: Kurt.Langfeld@uni-tuebingen.de \hfill } 

\centerline{ 
Talk presented at the International conference on {\it Strong and electro-weak
matter}, \hfill } 

\centerline{ October 2-5, 2002; Heidelberg, Germany \hfill }
}
\end{quote}
\eject

{\bf Introduction.} 
While the QCD phase diagram is well understood as function of the 
temperature $T$ for small baryon chemical potential $\mu $, very little 
is known about the phase diagram as function of $\mu $ and small 
temperatures. At very large densities, one might expect that a 
Fermi surface of quarks exists by virtue of asymptotic freedom. 
The residual weak gluonic interactions therefore lead to the 
formation of a condensate in certain di-quark 
channels~\cite{bai94,Rajagopal:2000wf}. 
For a baryon chemical potential of the order of several GeVs, 
the Fermi surfaces smears out due to the strong interactions, and 
when confinement effects become important at small $\mu $, the 
quark matter is re-arranged forming a Fermi surface of baryons. 
In my talk, I will address this transient regime. 

\vskip 0.2cm
{\bf The effective theory.} 
For this purpose, I will develop a lattice effective theory for 
the degrees of freedom which are relevant at the transition: 
the Yang-Mills degrees of freedom, and the composite scalar field 
representing the 2-flavor di-quark field, 
$\phi _A  = \epsilon _{ABC} \epsilon ^{ik} )(\bar{q}_c)^B_i \gamma _5 
q^C_k$, where $A,B,C=1 \ldots 3$ denote the color index of the 
SU(3) fundamental representation and $i,k$ represent the flavor 
index. $q$ and $q_c$ are the quark spinor and its charge
conjugated. I am assuming the following evolution of scales: 
even if one encounters a first order transition, one expects 
that the relevant correlation length grows when the chemical 
potential approaches the critical values from below, $\mu \rightarrow 
\mu _c^-$. In the low density phase, di-quarks are confined within the 
baryon. The crucial assumption is that e.g.~the electro-magnetic 
mean square radius increases less rapidly than the correlation 
length. In this case, one might treat the composite di-quark field 
as a point-like Higgs field belonging to the fundamental representation 
of the Yang-Mills color group. The effective theory is 
determined by gauge invariance and renormalizability. 
The Higgs Lagrangian with a non-vanishing di-quark chemical potential 
included is therefore given by
\bea 
L_{\mathrm Higgs} &=& [D_\mu \phi (x) ]^\dagger \; 
[D_\mu \phi (x) ] \; + \; \bigl[m^2 \, - \, \mu ^2 \bigr] 
\, \phi ^\dagger (x) \phi (x) 
\label{eq:5} \\ 
&+& \lambda \bigl[ \phi (x) ^\dagger \phi (x) \bigr]^2  
\; - \;  i \; \mu \; Q_0(x) \; , 
\nonumber 
\ena 
where 
\be
Q_0(x) \; = \; \frac{1}{i} \Bigl\{ \phi ^\dagger (x) 
D_0 \phi (x) \; - \; \bigl[D_0 \phi (x) \bigr]^\dagger \phi (x) 
\Bigr\}
\label{eq:6}
\en 
is proportional to the Baryon density, which, in the present 
effective theory, is entirely provided by the scalar currents. 
One observes that for $\mu \stackrel{>}{_\sim } m$ the theory 
is pushed towards the Higgs, i.e., the color-superconducting, phase. 
The aim of the present investigation is to study the competition 
between the confining forces and the di-quark correlations
of the superconducting phase. 

\vskip 0.2cm 
{\bf Bose-Einstein condensation versus confinement. }
To my knowledge, lattice Yang-Mills theory is the only possibility 
to put rigor to an effective theory designed for a description of 
confinement as well as condensation effects. In this first 
investigation, I will adopt a coarse graining point of view: 
I will study an $SU(2)$ gauge group with a fundamental Higgs field 
rather than the realistic gauge group $SU(3)$. I will drop the 
imaginary part of the action which gives rise to severe sign 
problem in practical simulations~\cite{Chandrasekharan:1999cm}. 
With the latter simplification, 
one looses the interpretation of $\mu $ as di-quark chemical potential. 
An inspection of (\ref{eq:5}), however, shows 
that this parameter can still be used to change the sign of the mass
squared term of the Landau theory and to drive the theory towards the 
Higgs phase. The complete lattice effective action, which is 
subject of the simulations below is given by 
\bea 
S_{\mathrm latt} &=& \beta \, \sum _{p} \; \frac{1}{2} \; \tr  \, U_p 
\label{eq:16} \\ 
&+& \kappa \; \sum _{x \, \nu} \; \rho _x \, \rho _{x+\nu}
\Bigl[ 1 + \bigl( \cosh (\mu a) -1 \bigr) \, \delta
_{\nu 0} \Bigr] \; \tr \bigl\{ 
\alpha (x) \; U_\nu (x) \; \alpha ^\dagger (x + \nu ) 
\; \bigr\} 
\nonumber \label{eq:17} \\ 
&-& \sum _{x} \Bigl\{ \lambda 
\bigl[ \rho ^2_x  - 1 \bigr]^2 \; + \; \rho ^2_x \Bigr\}
\nonumber 
\ena 
where $p$ denotes the plaquettes of the discretized space-time, and 
$U_p$ is the plaquette matrix constructed in the usual way from the 
link variables. $\kappa $ is the Higgs hopping parameter, and $\lambda
$ parameterizes the strength of the quartic Higgs coupling. 
The $SU(2)$ matrix $\alpha $ represents the angular part of 
the Higgs field, and $\rho (x)$ its modulus.

\begin{figure}[t]
\centerline{\epsfxsize=2.8in\epsfbox{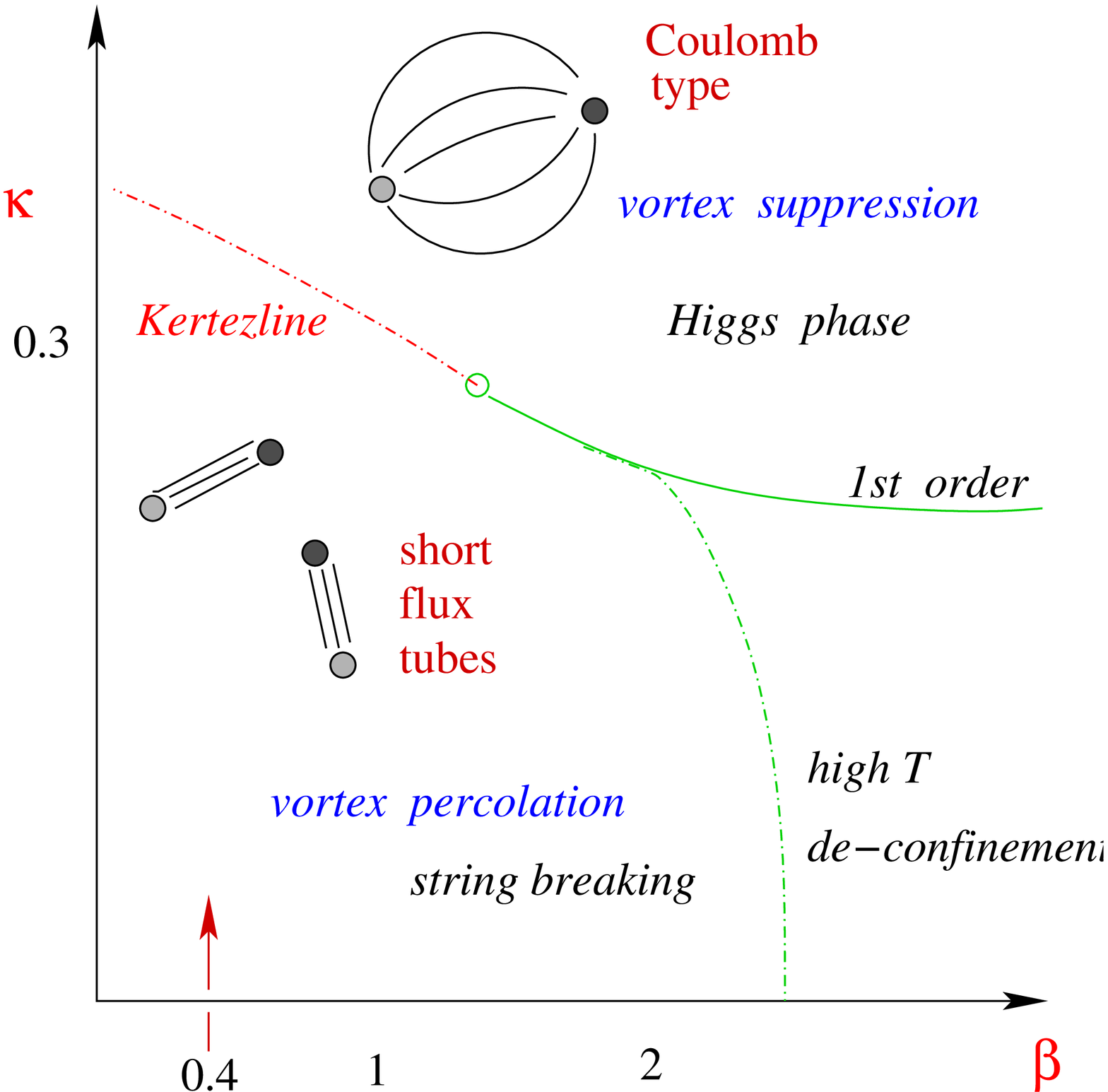} 
            \epsfxsize=2.8in\epsfbox{cross.eps}}   
\caption{The phase diagram of the SU(2) Yang-Mills theory with a 
fundamental Higgs field; schematic sketch (left panel); $E$ and the 
vortex density as function of $\kappa $ in the crossover regime. 
\label{fig:1}}
\end{figure}
\vskip 0.2cm
{\bf Litmus papers. } 
For a detection of the Higgs phase, one usually~\cite{Bunk:1992kf}
uses the expectation values 
of the spatial Higgs hopping term, i.e., $E = \langle \tr \phi ^\dagger _x 
U_{x,k} \phi _{x+k } \rangle $, $k\in \{1,2,3\}$. One observes that 
$E$ jumps when the first order transition line (see figure \ref{fig:1}) 
is crossed. Alternatively~\cite{Langfeld:2001he}, one might use that 
fact that, after 
gauge fixing, the residual global color symmetry is spontaneously 
broken in the Higgs phase. Here, I use 
Landau gauge for these purposes. If $\phi ^f$ denotes the gauge 
fixed Higgs field, I use $ 
\langle \phi ^f \rangle \; = \; \left\langle \left( 
\frac{1}{N} \sum _x \phi ^f(x) \right)^2 \right\rangle ^{1/2}  
$ as an indicator for color-superconductivity. 

\vskip 0.2cm 
The properties of the so-called center vortices are ideal candidates 
to understand quark confinement. Using the so-called Maximal 
Center Projection~\cite{DelDebbio:1996mh}, these vortex matter was
observed to extrapolate properly to continuum \break
theory~\cite{Langfeld:1997jx}. A percolating vortex cluster 
signals the formation of color-electric flux tubes: at zero density, this 
leads to quark confinement, while at non-zero densities 
short flux tubes are formed until the color-electric string 
breaks (see figure \ref{fig:1}). It will turn out that the 
(planar) vortex density is the interesting observable in the present 
case. 

\vskip 0.2cm 
In order to address the question whether quark- and gluon-deconfinement 
occurs at the same $\mu _c$, a litmus paper for the liberation 
of gluons is required. The vacuum energy density is the ideal 
quantity, since it rapidly rises when gluonic black body radiation 
becomes possible due to de-confinement. Here, I use the 
expectation value of the difference between the spatial and time-like 
plaquettes, which is know to contribute the major part to the energy 
density~\cite{Engels:1994xj}.

\begin{figure}[t]
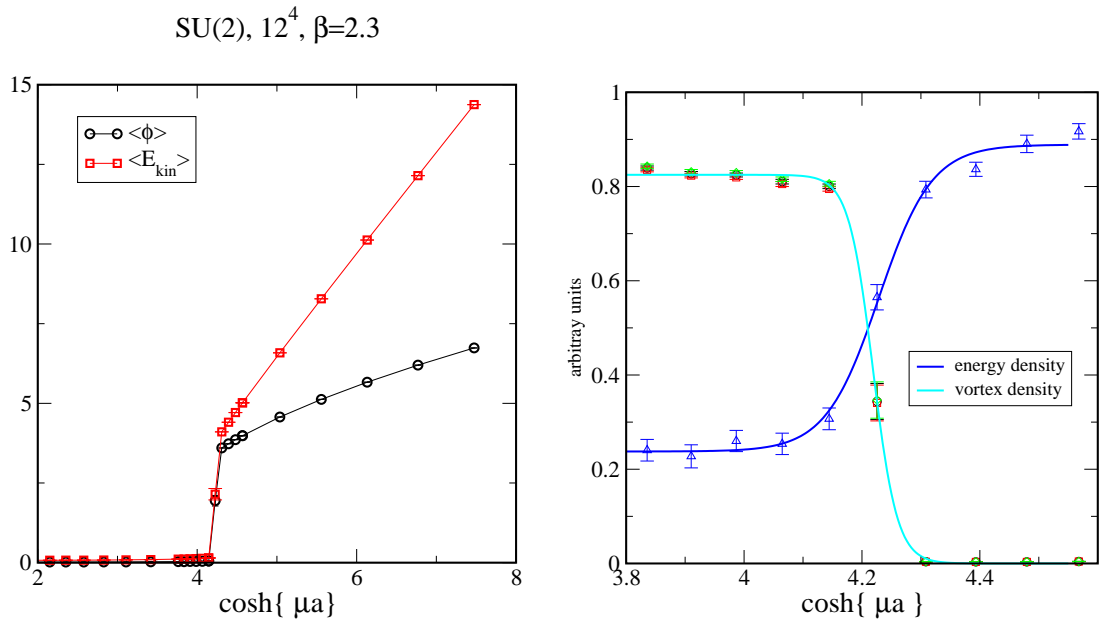

\centerline{\epsfxsize=2.7  in\epsfbox{phaset.eps} \hspace{.3cm}
            \epsfxsize=2.8 in\epsfbox{den.eps}}   
\caption{ The transition from the confinement to the 
color-superconducting phase 
\label{fig:2}}
\end{figure}
\vskip 0.2cm 
{\bf Results.} 
Let us first study the phase-diagram of the effective lattice theory 
as function of $\kappa $ and $\beta $ 
for a vanishing chemical potential. I find that the crossover at 
small values $\beta $ as function of $\kappa $ is accompanied by 
a drastic reduction of the vortex density. These results are in 
agreement with the recent achievements reported in~\cite{faber}. 
With the help of the vortex density, it is possible to extend the 
line of the phase transitions (green lines) into the crossover 
regime. Figure \ref{fig:1} 
(left panel) illustrates my findings concerning the phase diagram. 
The right panel shows the result of lattice simulation using 
a $6^4$ lattice, $\beta =0.4$. 
Hence, the line provided by the vortex suppression (red line in figure 
\ref{fig:1} left panel) is able to separate between the regime 
of short color electric flux tubes and the regime which has lost is 
capability to form color-electric strings. This line can be viewed as 
the analog of the ``Kert\'esz-line''~\cite{ker,Satz:2001zf} 
of the Ising model. The center vortices in the present case 
can be used in the same manner as the 
Polyakov lines in the context of Yang-Mills theory with 
dynamical quarks: in the latter case, the ``Kert\'esz-line'' of 
percolating Polyakov lines separates the phase with a broken 
color electric string from the Coulomb type
phase~\cite{Fortunato:2000ge}. 

\vskip 0.2cm 
Subsequently, I have chosen the parameters $\beta =2.3$, $\kappa =0.1$, 
$\lambda =0.01$, $12^4$ lattice to lie within the regime 
of the broken flux tubes. I have then increased the 
parameter $\mu $. At small values $\mu $, the confining effects 
dominate and the dependence of $E$ with $\mu $ is small. 
If $\mu $ exceeds a critical value, the onset of
color-superconductivity is signaled by the rapid increase of $E$ and 
a value of $\langle \phi ^f \rangle $ significant from zero 
(see figure \ref{fig:2} left panel). At the same time, the vortex
density rapidly drops indicating that the regime of short flux tubes
was abandoned. The drop of the vortex density is accompanied by 
the rise of the difference between the spatial and time-like 
plaquettes. This indicates that liberation of gluons and 
quarks takes place at the same critical value for $\mu $. 

\vskip 0.2cm
{\bf Acknowledgments: } Many helpful discussions with U.-J.~Wiese 
are greatly acknowledged. I thank S.~Hands for helpful comments. 
I am grateful to the organizers for the possibilty 
to present my results at this interesting meeting.

\end{document}